\documentclass[pre,twocolumn,showpacs,preprintnumbers,amsmath,amssymb]{revtex4}

\usepackage{graphicx}
\usepackage{dcolumn}
\usepackage{bm}

\begin{document}


\title{Multifractal Analysis of Human Retinal Vessels}

\author{Tatijana Sto\v si\' c}
\author{Borko D. Sto\v si\' c}
\email{borko@ufpe.br}
\affiliation{
Departamento de Estat\' \i sica e Inform\' atica, 
Universidade Federal Rural de Pernambuco,\\
Rua Dom Manoel de Medeiros s/n, Dois Irm\~ aos,
52171-900 Recife-PE, Brasil
}

\date{\today}

\begin{abstract}
In this work it is shown that vascular structures of the human
retina represent geometrical multifractals, characterized by a
hierarchy of exponents rather then a single fractal dimension. 
A number of retinal images from the STARE database 
(www.parl.clemson.edu/stare) are analyzed, corresponding to both 
normal and pathological states of the retina. 
In all studied cases a clearly multifractal behavior is observed, where
capacity dimension is always
found to be smaller then the information dimension, which is in turn
always smaller then the correlation dimension,
all the three being  
significantly lower then the DLA (Diffusion Limited Aggregation) fractal dimension.
We also observe a tendency of images corresponding to the pathological 
states of the retina
to have lower generalized 
dimensions and a shifted spectrum range, in comparison with the normal
cases.
\end{abstract}

\pacs{05.40.-a, 61.43.Hv, 87.57.-s, 87.57.Nk}


\maketitle

Over the past decade, there have been several attempts
\cite{family,mainster,landini,avakian, lakshi,masters} 
in the direction of employing
the fractal dimension as a measure for quantifying the ``state"  of
human retinal vessel structures (considered as geometrical objects),
with the expectation that such analysis may
contribute to automatic detection of pathological cases, and
therefore to computerization of the diagnostic process.
While this is certainly a valid question with possibly high impact on
real world diagnostic issues, there are some issues that should be
addressed before such investigations may prove useful for the standard 
clinical practice.
First, the fact that retinal
vessels represent ``finite size" realizations of a fractal
growth process, imposes questions about how exactly should one go about
measuring the fractal dimension of a particular instance 
(e.g. an electronic image of a retinal vessel structure, taken from a given
angle, with a given resolution and lightning conditions).
A related question is to what extent these calculations may correspond to the 
limiting fractal (which would have been attained if the growth 
process could have been extended to infinity),
or equivalently, to what degree they may be compared with 
calculations on other such finite instances. 
Although various issues related to these questions have already been addressed
(for a current review see e.g. \cite{masters}),
it seems that many of them remain open for further investigation.
Second, some of these works \cite{avakian,landini} 
address the point that the retinal vessels may have different properties 
in different regions, and do indeed find different characteristics 
depending on the locale of measurement, although the procedures
adopted in these works are 
only remotely related to established concepts of multifractality,
and the corresponding commonly accepted procedures for its measurement
(see e.g. \cite{feder, vicsek, vicsek2, vicsek3, chabra, chabra2} and references therein).

In this work we concentrate on the latter of the above issues, that is,
we show that the human retinal vessel structures are geometrical multifractals, 
characterized by a hierarchy of exponents rather then a single fractal dimension. 
We analyze a number of retinal images from the STARE database \cite{stare}, 
corresponding to both normal and pathological states of the retina. 
In all cases we find clearly
multifractal behavior. The capacity (or box counting) dimension is always
found to be smaller then the information (or Shannon) dimension, which is in turn
always smaller then the correlation dimension.
In all cases the observed values of the capacity dimension were  
significantly lower then the DLA (Diffusion Limited Aggregation) fractal dimension,
which has been considered in earlier works \cite{family,mainster,masters} as the primary model
relevant for the phenomenon at hand.
It is also found that images corresponding to pathological cases 
tend to have lower generalized 
dimensions, as well as a shifted spectrum range, 
in comparison with the normal cases.

In contrast to regular fractals (or monofractals), 
multifractals are characterized by a
hierarchy of exponents, rather then a single fractal dimension.
A well known example of multifractality is the
growth probability distribution during the DLA growth process,
which has been shown to exhibit multifractal scaling 
\cite{amitrano,hayakawa,nittmann, ohta}.
Geometrical (or mass) multifractals represent a special case
when the measure of interest is homogeneously distributed over
the observed structure, so that only the number of particles (Lebesgue measure)
contributes to the measure within a given region of the fractal \cite{vicsek,vicsek2}.
Considering a structure with mass (number of pixels) $M_0$ and linear
size $L$, covered with a grid of boxes of linear size $\ell$,
the generalized dimension $D_q$ for the mass distribution is defined by
\begin{equation}
\sum_i\left({\frac {M_i}{M_0}}\right)^q\sim\left({\frac {\ell}{L}}\right)^{(q-1)D_q},
\label{one}
\end{equation}
where $M_i$ is the mass (number of pixels) within the $i$-th box, 
and $q$ is a continuous (adjustable) variable that makes it possible 
to single out fractal properties of the object at different scales
(equivalent of inverse temperature in thermodynamics). 
The generalized dimensions $D_0$, $D_1$ and $D_2$
correspond to the capacity (or box-counting) dimension, 
information (or Shannon) dimension,
and correlation dimension, respectively.
Finally, $D_{-\infty}$ and $D_{\infty}$ represent the limits of the
generalized dimension spectrum (for monofractals, all the generalized
dimensions coincide, being equal to the unique 
fractal dimension).

It turns out that the direct application of (\ref{one}) in practice
is hindered by the fact that for $q < 0$ the boxes that contain a small
number of particles (because they barely overlap with the cluster)
give anomalously large contribution to the sum on the left hand side
of (\ref{one}).
To alleviate this problem and 
perform the multifractal analysis of the retinal vessel structures,
we adopt the generalized sand box method \cite{vicsek2,vicsek3},
which has been successfully used do demonstrate geometric
multifractality of DLA \cite{vicsek2}.
This procedure consists in randomly selecting $N$ points
belonging to the structure, and counting for each such point $i$ the
number of pixels $M_i(R)$ that belong to the structure, inside boxes
of growing linear dimension $R$,  centered at the selected pixels.
The left hand side of equation (\ref{one}) can now be interpreted as 
the average of the quantity $\left(M_i(R)/M_0\right)^{q-1}$ according to
probability distribution $M_i(R)/M_0$. When the box
centers are chosen randomly, the averaging should be made over the 
chosen set, and the equivalent of (\ref{one}) becomes
\begin{equation}
\left<\left({\frac {M(R)}{M_0}}\right)^{q-1}\right>\sim
\left({\frac {R}{L}}\right)^{(q-1)D_q}.
\label{two}
\end{equation}
The practical advantage of this method is that the boxes are centered 
on the structure, so that by construction there are no boxes with too few particles 
(pixels) inside.

To verify whether human retinal vessel structures 
demonstrate geometrical multifractal scaling properties,
we have used a set of forty retinal images 
from the STARE database \cite{stare},
manually segmented by two different observers 
(herefrom referred to by initials AH and VK as in \cite{stare})
from twenty original retinal scans 
(containing ten normal and ten pathological cases),
for the purpose of studies on automatic image segmentation 
and diagnostics \cite{stare2}.
The images segmented by observers AH and VK differ
in level of detail, and the resulting set, 
totaling forty segmented images, is available for download from
the STARE project \cite{stare}
in ppm file format,
with resolution of 700x605 pixels.
As recently it has been argued \cite{parsons} 
that fractal analysis may be more sensitive to changes in
vascular patterns when skeletal images of vascular trees are considered, rather then
the original segmented images (which contain the vessel width information),
in order to verify whether the vessel width information indeed does exert 
influence on the multifractal analysis, 
we have also performed skeletonization of the two downloaded sets 
using the eight connected Rosenfeld algorithm \cite{rosenfeld}, to produce two
additional sets of twenty images each. 
A typical normal and a pathological image,
segmented by observers AH and VK 
(where images segmented by observer VK demonstrate a substantially higher level of
detail), respectively, together with their skeletonized versions using the Rosenfeld algorithm,
are shown in Fig.~1.
\begin{figure}
\includegraphics[width=3.2in]{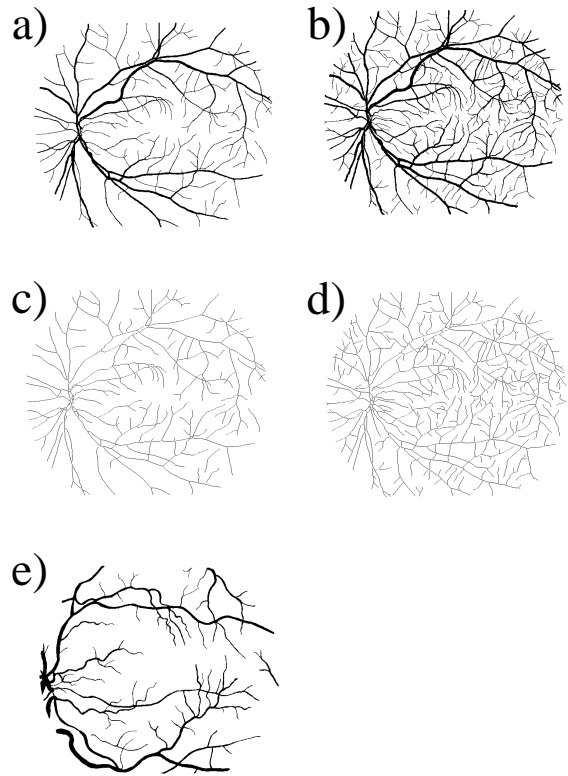}
\caption{\label{fig1}
Image of a typical normal retinal vessel structure
(image files im0162.ah.ppm from the STARE database \cite{stare}),
segmented by a) observer AH and b) observer VK, together with their
skeletonized versions c) and d), respectively, and a typical
pathological structure 
(image file im0001.ah.ppm, diagnosed with Background Diabetic Retinopathy),
segmented by observer AH.
}
\end{figure}

For all of the four sets (containing twenty images each), we have performed measurements 
(calculations) according to (\ref{two}), selecting 1000 random points on each structure
(typical structure size $M_0$ is approximately 30000 pixels,
and the typical linear size $L$ is 600 pixels),
and counting number $M_i$ of pixels inside boxes centered at selected points.
These numbers were then used to extract
generalized dimension $D_q$, for different values of q ($-10<q<10$), as slopes
of the lines obtained through regression (minimum squares fitting) of 
${\log{\left<{\left[{M(R)/M_0}\right]}^{q-1}\right>}}/(q-1)$, 
as a function of $\log(R/L)$.
The whole procedure was repeated 100 times 
(with different random choices of the 1000
random points), for each image, and for each value of $q$.
The final values of $D_q$ were calculated as averages
over these repetitions.

A word is due on calculations for the special case $q=1$, corresponding
to information dimension $D_1$. As the above formulas are non-analytic for 
$q=1$, we perform calculations at $q=1\pm \epsilon$, for $\epsilon=0.001$, and assuming
linearity of the function $D(q)$ in this (short) interval, we interpolate
$D_1\approx\left(D_{1-\epsilon}+D_{1+\epsilon}\right)/2$
(in fact, the difference between the values of $D_q$ calculated on both 
sides of $q=1$ was found to be only slightly larger then
the statistical fluctuations induced by
random choice of the set of measurement points on the structure).

\begin{figure}
\includegraphics[width=3.2in]{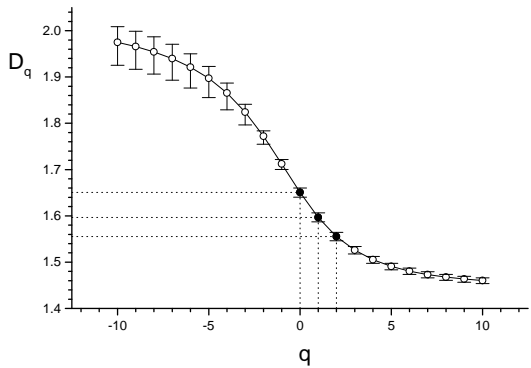}
\caption{\label{fig2} 
The generalized dimension spectrum, $D_q$ versus $q$,
for a typical normal retinal image 
(image file im0162.ah.ppm \cite{stare}), averaged over 100 random choices
of 1000 points each (see text for details). 
The error bars indicate the largest and smallest values  encountered within
the 100 runs, and the curve connecting the points serves as a guide to the eye.
The points corresponding to the capacity dimension $D_0=1.647$,
the information dimension $D_1=1.594$ and the correlation dimension
$D_2=1.552$ are represented by full circles, while the dotted lines
serve to emphasize their position.
}
\end{figure}

Results of a typical calculation are shown in Fig.~2, where it is seen that
the observed retinal vessel structure clearly demonstrates multifractal 
scaling, rather then being a simple monofractal 
(there is a significant difference between generalized dimensions).
In particular, the capacity dimension $D_0$,
the information dimension $D_1$ and the correlation dimension $D_2$
are all different, satisfying $D_0>D_1>D_2$.
Also, all the three values remain substantially lower then the 
DLA fractal dimension estimate, commonly
accepted in the literature, of
$D_{q=2}\simeq 1.71$ (which is in fact underestimated
by commonly used methods) \cite{vicsek2},
in contrast with previous findings \cite{family,mainster,masters}.

Numerical corresonding to Fig.~2 
(for the set of twenty images from the STARE database segmented by observer AH)
are given in Tab.~\ref{tab1}. 
The first column lists the image names, while the second column
indicates image classification status as 
``Pathological" or ``Normal"
\cite{stare3}.
The values of generalized dimensions $D_q$ are given for $q=-10,0,1,2,10$,
where as already mentioned $D_0$, $D_1$ and $D_2$ correspond to the 
capacity, information and correlation dimension, respectively, 
while $D_{-10}$ and
$D_{10}$ indicate the range of the general dimension spectrum.
\begin{table}
\caption{
Generalized dimensions $D_q$ for $q=-10,0,1,2,10$,
for the twenty analyzed images from the STARE database.
The second column indicates classification status for
each of the images (pathological and normal).
}
\label{tab1}
\begin{tabular}{lclllll}
Image&Status&$D_{-10}$&$D_0$&$D_1$&$D_2$&$D_{10}$\\
\hline
im0001.ah&P&1.968&1.540&1.494&1.462&1.361\\
im0002.ah&P&1.930&1.548&1.498&1.460&1.370\\
im0003.ah&P&1.877&1.509&1.469&1.443&1.380\\
im0004.ah&P&1.777&1.522&1.492&1.465&1.367\\
im0005.ah&P&1.900&1.589&1.560&1.538&1.474\\
im0044.ah&P&1.886&1.541&1.493&1.459&1.363\\
im0077.ah&N&1.911&1.576&1.528&1.496&1.426\\
im0081.ah&N&1.917&1.555&1.514&1.487&1.421\\
im0082.ah&N&1.981&1.578&1.518&1.476&1.404\\
im0139.ah&P&1.904&1.565&1.516&1.481&1.413\\
im0162.ah&N&1.968&1.647&1.594&1.552&1.459\\
im0163.ah&N&1.998&1.642&1.587&1.550&1.476\\
im0235.ah&N&1.945&1.597&1.548&1.514&1.442\\
im0236.ah&N&1.868&1.584&1.544&1.514&1.448\\
im0239.ah&N&1.945&1.587&1.549&1.520&1.437\\
im0240.ah&N&1.918&1.593&1.564&1.543&1.494\\
im0255.ah&N&1.944&1.633&1.604&1.583&1.521\\
im0291.ah&P&1.819&1.516&1.482&1.454&1.348\\
im0319.ah&P&1.703&1.443&1.409&1.382&1.299\\
im0324.ah&P&1.923&1.567&1.520&1.486&1.399\\
\end{tabular}
\end{table}
It is seen from Tab.~\ref{tab1} that all of the values calculated
for the capacity dimension (which corresponds to box counting method),
and indeed the correlation dimension (corresponding to methods
such as radius of gyration or the density-density correlation function),
are significantly lower then the DLA fractal dimension
$D_{q=2}\simeq 1.71$ \cite{vicsek2}.
Therefore, our results show that retinal vessel structures are
geometrical multifractals, and that the overall fractal dimension
is lower then that of the DLA.

Results of the multifractal analysis for the other three sets
of images (STARE database images segmented by observer VK, 
and the skeletonized versions of AH and VK) all yield
qualitatively similar results,
all of them clearly demonstrating multifractal behavior.
In Table \ref{tab2} we present the results for the capacity 
(box counting) dimension $D_0$,
for all of the four sets of images.
It follows from Table \ref{tab2} that the process of skeletonization
(removal of vessel width information from the image) slightly reduces
the fractal dimension, while this effect is much weaker in comparison with the
effect of the level of detail present in the segmentation process, as found
between the two current observers. 
However, when the results are compared consistently within each group separately,
the mean fractal dimension is found to be lower
for the pathological images then for the normal cases, for all of the four groups.
Although this finding can hardly be considered conclusive from the statistical viewpoint,
it is nevertheless encouraging from the point of view that fractal spectrum analysis 
could be employed for quantification of the retinal vessel state,
in order to contribute to automatic diagnostics. 
To this end, far more detailed studies of images 
corresponding to specific deseases and normal cases, are needed.
Assuming that each of the observers
consistently applied his own criteria in segmentation, it follows that
the fractal dimension results may be compared only between images
segmented by the same observer, either skeletonized or not, but should be
normalized before making comparisons of results from different groups.
\begin{table}
\caption{\label{tab2}
Capacity(or box counting) dimension $D_0$ for the two sets
of images from the STARE database segmented by observers AH and VK,
together with their skeletonzied versions.
The second column indicates classification status for
each of the images (pathological and normal),
and the last three lines present averages for the
pathological, normal and all images, respectively.
}
\begin{tabular}{lcllll}
Image&Status&AH&AH-S&VK&VK-S\\
\hline
im0001&P&1.540&1.545&1.583&1.593\\
im0002&P&1.548&1.524&1.574&1.568\\
im0003&P&1.509&1.500&1.593&1.608\\
im0004&P&1.522&1.508&1.573&1.598\\
im0005&P&1.589&1.554&1.680&1.663\\
im0044&P&1.541&1.538&1.668&1.661\\
im0077&N&1.576&1.591&1.658&1.662\\
im0081&N&1.555&1.551&1.668&1.671\\
im0082&N&1.578&1.585&1.665&1.680\\
im0139&P&1.565&1.564&1.679&1.678\\
im0162&N&1.647&1.638&1.714&1.700\\
im0163&N&1.642&1.612&1.684&1.646\\
im0235&N&1.597&1.588&1.685&1.675\\
im0236&N&1.584&1.581&1.658&1.662\\
im0239&N&1.587&1.597&1.655&1.655\\
im0240&N&1.593&1.563&1.677&1.663\\
im0255&N&1.633&1.634&1.696&1.693\\
im0291&P&1.516&1.491&1.604&1.578\\
im0319&P&1.443&1.446&1.555&1.561\\
im0324&P&1.567&1.503&1.642&1.617\\
\hline
Average&P&1.534&1.517&1.615&1.612\\
&N&1.599&1.594&1.676&1.671\\
&All&1.567&1.556&1.646&1.642\\

\end{tabular}

\end{table}

When addressing multifractality, numerous works deal with the 
so-called $f(\alpha)$ spectrum (see e.g. \cite{halsey,feder, vicsek} and references therein), where 
\begin{equation}
N(\alpha)=L^{-f(\alpha)} ,
\label{three}
\end{equation}
represents the number of boxes $N(\alpha)$ where the probability
 $P_i$ of finding a particle (pixel) within a given region $i$ scales as
\begin{equation}
P_i=L^{\alpha_i} ,
\label{four}
\end{equation}
and $f(\alpha)$ may be understood as the fractal dimension of the union of regions
with singularity strenghts between $\alpha$ and $\alpha+d\alpha$.
The exponent $\alpha$ takes values from the interval $[-\infty,\infty]$,
and the function $f(\alpha)$  is usually a single humped function
with a maximum at $df(\alpha(q))/d\alpha(q)=0$. The relationship
between the $D(q)$ spectrum and the $f(\alpha)$ spectrum is made via the Legendre
transform
\begin{equation}
f\left(\alpha\left(q\right)\right)=q\alpha\left(q\right)-\tau\left(q\right),
\label{five}
\end{equation}
where
\begin{equation}
\alpha\left(q\right)={\frac {d\tau(q)}{dq}},
\label{six}
\end{equation}
and
\begin{equation}
\tau(q)\equiv (q-1)D_q
\label{seven}
\end{equation}
is the mass correlation exponent of the $q^{th}$ order.
To calculate the derivatives in (\ref{six}), we have
performed calculations at pairs of points $q$ and $q+\epsilon$ with
$\epsilon=0.001$, so that derivatives were calculated as 
$d\tau(q)/dq\approx(\tau(q+\epsilon)-\tau(q))/\epsilon$, except at point $q=1$,
where we have used $d\tau(q)/dq\approx(\tau(1+\epsilon)-\tau(1-\epsilon))/(2\epsilon)$.
\begin{figure}
\includegraphics[width=3.2in]{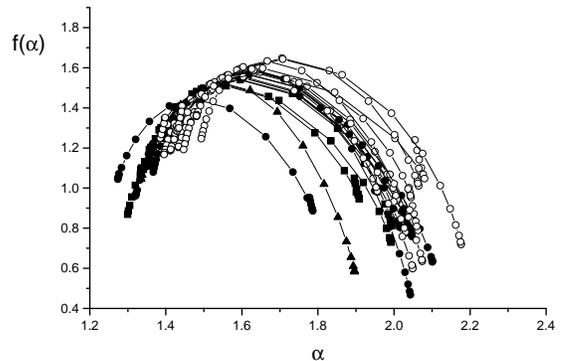}
\caption{\label{fig3} 
The $f(\alpha)$ spectrum for the twenty images 
from the STARE database \cite{stare}, segmented by observer AH.
Curves corresponding to normal retinal images
are represented by open circles, and those corresponding
to pathological images \cite{stare3}
are represented by full symbols.
It is seen that pathological image curves tend to be shifted
to the lower $\alpha$ range and have lower maxima, 
in comparison with the normal images 
(see text for more details).
}
\end{figure}

In Fig.~\ref{fig3} we show detailed results of our calculations, performed
on the STARE database images segmented by observer AH,
with respect to the $f(\alpha)$ spectrum.
While the current set of images is not particularly adequate for testing the
effects of a given type of pathology (there are only ten normal images,
and ten pathological images affected by not necessarily the same disease,
see \cite{stare3}),
it is seen that pathological case images tend to have 
lower maxima, occasionally more narrow spectrum range,
and a shift in the spectrum position,
in comparison with the normal cases.
 
\begin{figure}
\includegraphics[width=3.2in]{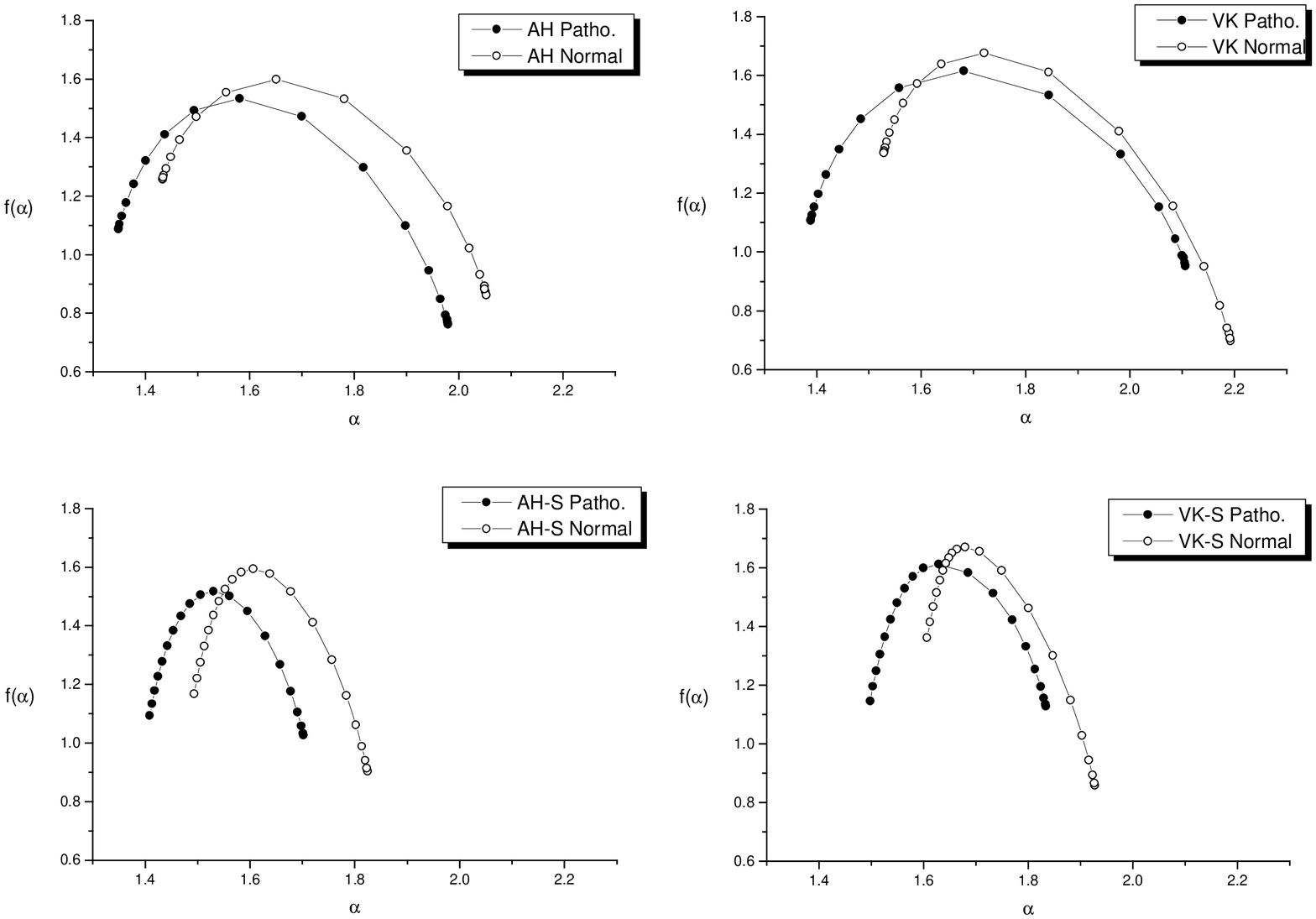}
\caption{\label{fig4} 
The $f(\alpha)$ spectrum for the twenty images 
from the STARE database \cite{stare}, segmented by observer AH.
Curves corresponding to normal retinal images
are represented by open circles, and those corresponding
to pathological images \cite{stare3}
are represented by full symbols.
It is seen that pathological image curves tend to be shifted
to the lower $\alpha$ range and have lower maxima, 
in comparison with the normal images 
(see text for more details).
}
\end{figure}

Finally, in Fig.~\ref{fig4} we present results of the $f(\alpha)$ spectrum
averaged separately for the normal and the pathological images for all of the four
sets, where it is seen that the previous observation holds for both
observers, independent of skeletonization. The skeletonized images present
more narrow $f(\alpha)$ spectrum then the original segmented images 
(which contain the vessel width information) for both observers,
which may explain the conclusion of
\cite{parsons} that fractal analysis after skeletonization
may be more sensitive to changes in
vascular patterns. More precisely, since monofractals have infinitely narrow
$f(\alpha)$ spectrum (a single fractal dimension),
the above results show that skeletonized structures may be more closely 
approximated as monofractals (when a single dimension is calculated rather then
the whole spectrum). As the general properties of the spectrum
are preserved through skeletonizatin, another advantage of using such images
may be considered the fact that they contain far fewer pixels, and therefore
the calculations require less computer time.

The results of calculations of the $f(\alpha)$ spectrum
presented in Figs.~\ref{fig3}-\ref{fig4} 
again may be considered encouraging from the point of view of
the objective of turning the diagnostic process automatic, although
further more detailed studies are necessary to determine their 
statistical significance, and whether the observed
differences in multifractal scaling behavior may be exploited for
discerning normal images from images with certain types of pathologies.
More precisely, the current work is primarily concerned with establishing
the fact that retinal vessel images represent geometrical multifractals,
nevertheless, our calculations suggest that there may be grounds for 
automatic differentiating
between normal images and certain pathological cases.

In conclusion, we show in this work that vascular structures of the human
retina represent geometrical multifractals, characterized by a
hierarchy of exponents, rather then a single fractal dimension. 
We analyze twenty retinal images from
the STARE database \cite{stare}, where half of the images correspond to 
normal states of the retina, and half to different pathological states \cite{stare3},
together with their skeletonized versions.
In all studied cases we find clearly multifractal behavior, with capacity
dimension considerably lower then the DLA value.
We also observe a tendency of normal images of having higher generalized 
dimensions and a shift of the $f(\alpha)$ spectrum range towards higher 
singularity strength values, in comparison with the pathological
cases. While the last observations are hardly conclusive from a statistical
standpoint, they may prove relevant in the quest of automatic diagnostic
procedures.
\vfill


\end{document}